\documentclass[aps,prl,reprint,superscriptaddress]{revtex4-1}
\usepackage{graphicx}
\usepackage{rotating}
\usepackage{amssymb}    
\usepackage{amsmath}   
\usepackage{epsfig}
\usepackage[normalem]{ulem}   
\usepackage{bm}   
\usepackage{color}

\date{\today}

\begin{document}

\title{Real-time vulnerability of synchronized states}

\author{Everton S. Medeiros}
\email{esm@if.usp.br}
\affiliation{Institute of Physics, University of S\~ao Paulo, Rua do Mat\~ao, Travessa R 187, 05508-090, S\~ao Paulo, Brazil}
\affiliation{Institute for Chemistry and Biology of the Marine Environment, Carl von Ossietzky University of Oldenburg, Oldenburg, Germany}
\author{Rene O. Medrano-T}
\affiliation{Departamento de F\'isica, Universidade Federal de S\~ao Paulo,  Diadema, S\~ao Paulo, Brazil}
\author{Iber\^e L. Caldas}
\affiliation{Institute of Physics, University of S\~ao Paulo, Rua do Mat\~ao, Travessa R 187, 05508-090, S\~ao Paulo, Brazil}
\author{Tam\'as T\'el}
\affiliation{Institute for Theoretical Physics, E\"otv\"os Lor\'and University, P\'azm\'any P\'eter S\'et\'any 1/A, H-1117 Budapest, Hungary}
\affiliation{MTA-ELTE Theoretical Physics Research Group, P\'azm\'any P\'eter S\'et\'any 1/A, H-1117 Budapest, Hungary}
\author{Ulrike Feudel}
\affiliation{Institute for Chemistry and Biology of the Marine Environment, Carl von Ossietzky University of Oldenburg, Oldenburg, Germany}

\email{esm@if.usp.br}
\begin{abstract}
 The time-dependent vulnerability of synchronized states is shown for a complex network composed of electronic circuits. We demonstrate that disturbances to the local dynamics of network units can produce different outcomes to synchronization depending on the event timing. We address such time dependence by systematically perturbing the synchronized system at instants of time equally distributed along its trajectory. We find the time instants at which the perturbation desynchronizes the network to be complicatedly mixed with the ones that restore synchronization. Additionally, we characterize perturbation sets obtained for consecutive instants of time by defining a safety index between them. Finally, we demonstrate that the vulnerability is due to state space sensitivities occurring along synchronized trajectories. 
\end{abstract}

\maketitle

Many complex systems, ranging from technological devices to ecology and human physiology, are composed of smaller parts operating in synchrony in order to perform their global behavior \cite{Kurths2003}. For example, in power grids, the power generators have to remain synchronized to guarantee the frequency stability of the network. Failures in this state can lead to severe power outages \cite{Motter2013}. In ecology, phenological synchronization establishes the temporal overlap between interacting species. Such synchrony is now threatened by climate change \cite{Deacy2017}. In the heart, asynchronous pumping of the left and right ventricles leads to out of sync heart contractions causing severe health conditions due to low blood flow to the body \cite{Gray1998}.  

In networks, the asymptotic stability of synchronous states with respect to small perturbations is well determined in the linear limit by the formalism of the {\it master stability function} \cite{Pecora1998,Pecora2000}. Yet, the impact of large perturbations has been addressed only by measurements performed in the synchronization basins, i.e., the state space configuration in the initial instant \cite{Girvan2006,Menck2013}. Additionally, the fractality of the boundaries of such synchronization basins has been also identified as a source of the sensitivity of synchronized states to perturbations \cite{Medeiros2018}. 

However, instead of estimating whether a perturbation of an initial condition leads a network to desynchronization, one can also raise a different question. Suppose the system has already reached a completely synchronized state performing some oscillatory dynamics. One now asks how vulnerable is this synchronized state with respect to prescribed perturbations occurring at a certain moment of its dynamics? Does it matter at which time instant this perturbation occurs along the trajectory? The answers to these questions are essential for the safety of technological applications as well as for designing responsible interventions in natural systems.

To address this demand, we consider a random network composed of identical electronic circuits. By perturbing the synchronized state in chronological instants of time, we demonstrate that the susceptibility of synchronization to disturbances changes in a non-trivial manner along the system's trajectory. A perturbation applied at one time instant could lead to a restoration of the synchronized state, while the same perturbation applied in the next time instant, very close to the previous one, could desynchronize the whole network. We call this phenomenon real-time vulnerability of synchronized states. We analyze it by applying sets of perturbations in subsequent time instants. For these perturbation sets, we identify safe sets that still lead the network to synchronization and characterize their transformations by measuring the safety index, a measure of their alikeness. A basin stability analysis shows that the relative size of the safe sets does not change significantly between the consecutive time instants, suggesting that only the location of the sets is important. Finally, we attribute the phenomenon to the existence of an unstable chaotic set in the state space and show the mechanism at which this set influences the network.
 
We study a random network composed of $N$ electronic circuits with the dimensionless dynamics given by \cite{Stoop2010}: 

\begin{eqnarray}
      \label{Eq1}
        \nonumber
    \dot{x}_{i}  &=& \alpha x_i + z_i + \frac{\sigma}{D_i}\sum_{j\in D_i}(x_{j}-x_{i}), \\
      \dot{y}_{i}  &=& z_i - f(y_i) + \frac{\sigma}{D_i}\sum_{j\in D_i}(y_{j}-y_{i}), \\
      \dot{z}_{i}  &=& -x_i -\beta y_i +\frac{\sigma}{D_i}\sum_{j\in D_i}(z_{j}-z_{i}),
      \nonumber
          \end{eqnarray}
where $f(y_i)=\frac{\gamma}{2}(|y_i+\gamma^{-1}|-|y_i-\gamma^{-1}|)$ describes a piece-wise linear diode resistance with slope $\gamma$. The vector ${\bf v}_i(t)=(x_{i}(t),y_{i}(t),z_{i}(t))^{\top}$ defines the state space of each circuit $i$ with $i=1,...,N$. The parameters $\alpha$ and $\beta$ are related to circuit elements. Following Ref. \cite{Stoop2010}, for each circuit we fix $\alpha=0.6$, $\beta=2.18$, and $\gamma=470$. For this parameter set, the electronic circuit exhibits a stable limit cycle ${\bf A}$ with dimensionless period $T=16$, an unstable chaotic set, i.e., a chaotic saddle $\Lambda$, embedded in the basin of attraction of ${\bf A}$, and an attractor at infinity. The parameters $\sigma$ and $N$ are the coupling strength and the network size, respectively, fixed at $\sigma=0.1$ and $N=25$. The parameter $D_i$ specifies the number of units to which the circuit $i$ is connected. For informations on the structure of the network see Supplemental at Material \cite{Note1}.

Initially, all network circuits are set to synchronize at their limit cycle attractor ${\bf A}$. The synchronized state lies in a synchronization manifold ${\bf S}$, defined as ${\bf v}_1(t)={\bf v}_2(t)= \dots ={\bf v}_N(t)$, where all states, in the limit cycle, are generally denoted by ${\bf v_S}(t)=( x_S(t), y_S(t), z_S(t))^{\top}$. In order to distinguish the attractor $\mathbf{A}$ of a single uncoupled circuit from the synchronized oscillating dynamics of the high-dimensional system, we refer to the second as $\mathbf{A_S}$, the limit cycle attractor in the $3N$-dimensional state space. Additionally, in the $3N$-dimensional state space, we refer to the chaotic set as $\mathbf{\Lambda}$. The synchronized behavior of the network can be assessed by a next neighbour error, $E_i(t)=\|{\bf v}_i(t)-{\bf v}_{i-1}(t)\|$. The perturbation applied to one unit residing on ${\bf A_S}$ consists of a deviation ${\bf \Delta}_i=( \Delta x_i, \Delta y_i, \Delta z_i)^{\top}$, directly applied to the dynamical state of a preselected circuit, $i$. In this manner, its dynamical state right after a perturbation at a time instant $t_p$ is given by ${\bf v}_i(t_p)=(x_S(t_p)+\Delta x_i, y_S(t_p)+\Delta y_i,z_S(t_p)+\Delta z_i)^{\top}$. 

\begin{figure}[!htp] 
\centering
\includegraphics[width=8.7cm,height=4.2cm]{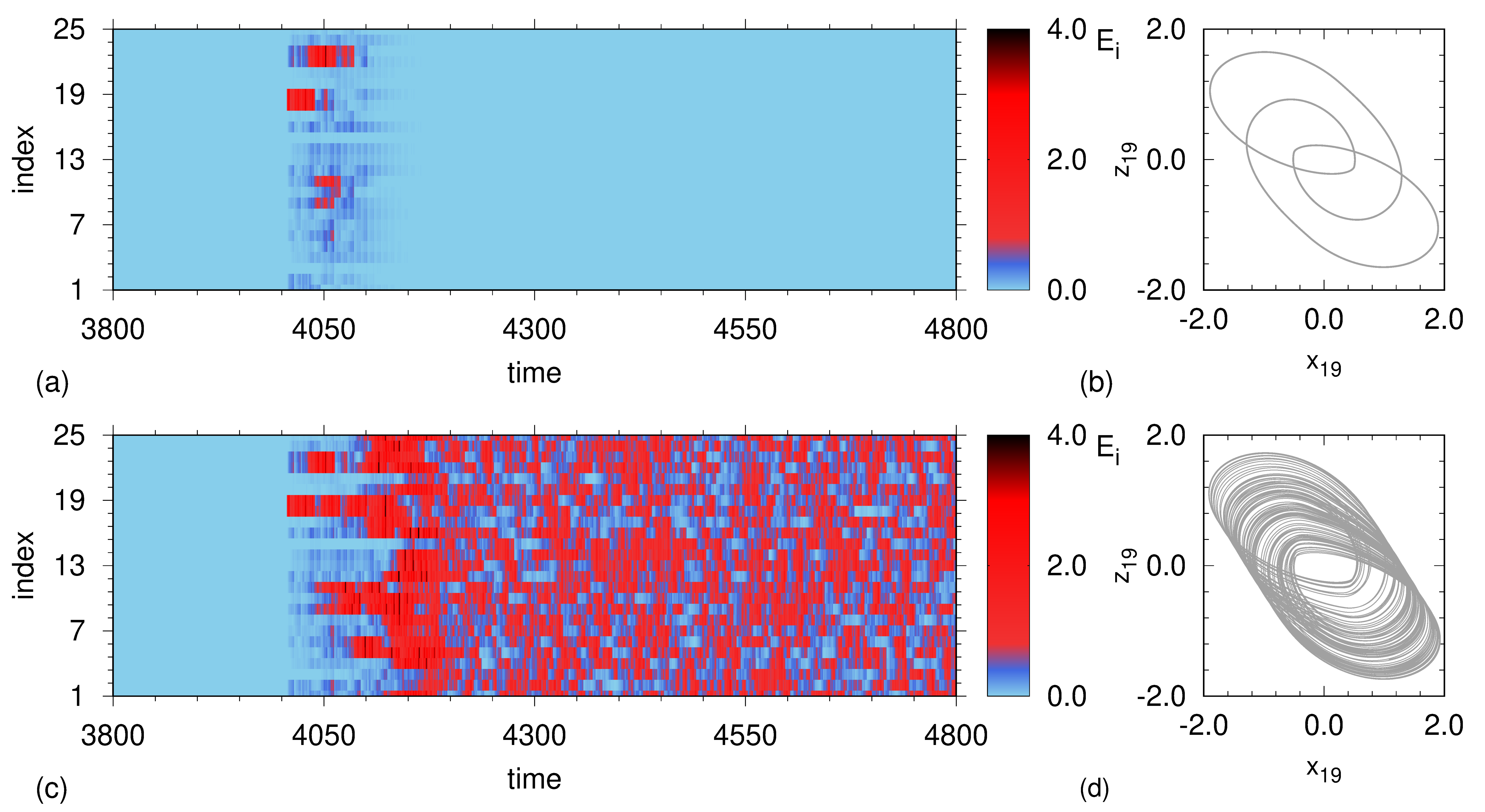}
\caption{Time evolution of the system. The color code indicates neighbour error, $E_i$. (a) The perturbation, ${\bf \Delta}_{19}=(0.0,0.97,-1.14)^{\top}$, to the local dynamics of the node $19$ is realized at the time $t_{p1}=4005.97$. (b) Synchronized oscillation. (c) Same perturbation applied to the node $19$ at the time $t_{p1}+\Delta t$ with $\Delta t=10^{-2}$. (d) Chaotic behavior observed in the state space of node $19$.}
\label{figure_1}
\end{figure}

The perturbation time instant $t_p$ is the central point of this work. In Fig. \ref{figure_1}(a), the network is perturbed at an arbitrary time instant $t_{p1}=4005.97$ by applying the perturbation ${\bf \Delta}_{19}=(0.0,0.97,-1.14)^{\top}$ to an arbitrary unit (say, $i=19$ with $D_{19}=10$). For an alternative perturbed unit see Supplemental Material at \cite{Note1}. We observe that the system returns to the synchronization manifold ${\bf S}$ leading to the conclusion that the perturbed state ${\bf v}_{19}(t_{p1})$ belongs to the synchronization basin $\mathcal{B}$ of ${\bf A_S}$, i.e. ${\bf v}_{19}(t_{p1}) \in \mathcal{B}({\bf A_S})$. As a consequence, the synchronization is restored after the perturbation and all units follow the same limit cycle $\mathbf{A}$ depicted in Fig. \ref{figure_1}(b). Now, if the same perturbation is applied at a slightly different instant of time, $t_{p1}+\Delta t$ with $\Delta t=10^{-2}$, we find that the network desynchronizes, indicating the opposite as before, i.e., ${\bf v}_{19}(t_{p1}+\Delta t) \notin \mathcal{B}({\bf A_S})$, as shown in Fig. \ref{figure_1}(c). Each unit is trapped in irregular trajectories (cf. Fig. \ref{figure_1}(d)). This disagreement suggests that the synchronization stability depends crucially on the particular instant of time at which a certain perturbation is applied. Consequently, the very same perturbation imposed to the system at the same circuit may not lead the network to normal functioning in synchrony.

\begin{figure}[!htp] 
\centering
\includegraphics[width=8.7cm,height=4.2cm]{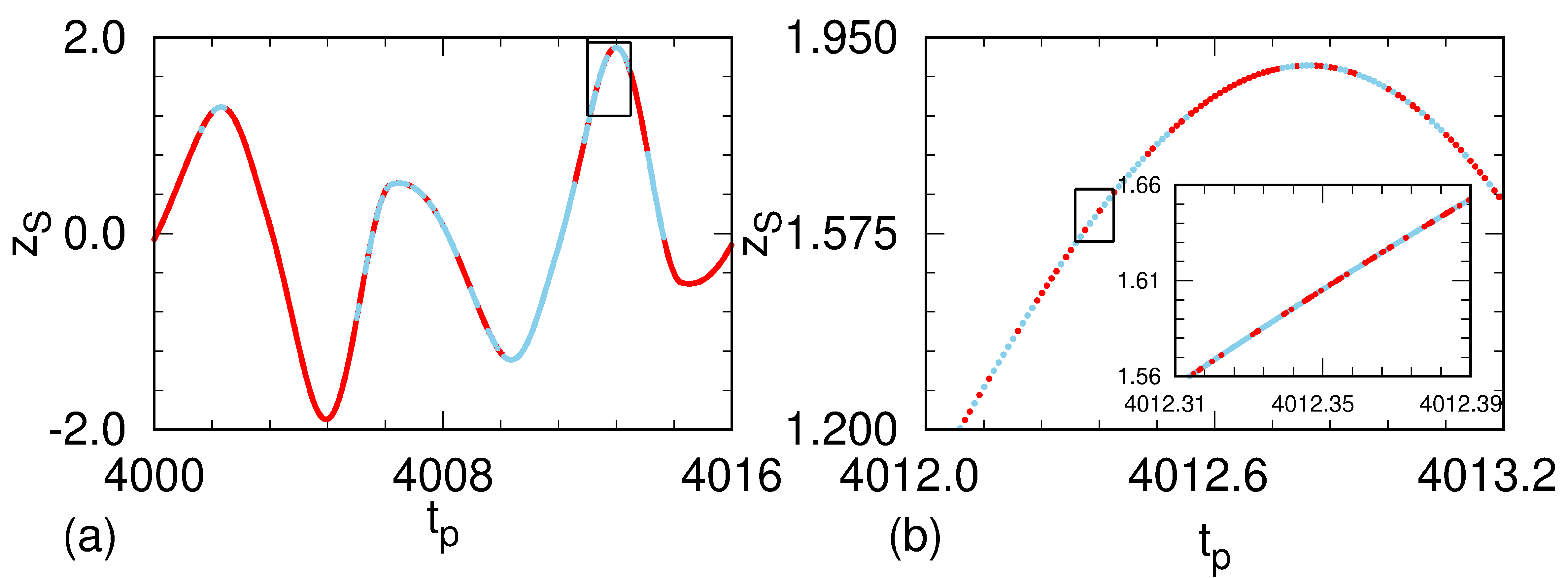}
\caption{(a) The $z$ coordinate of the synchronized oscillation as function of the perturbation time $t_p$. In red are the time intervals for which a perturbation, ${\bf \Delta}_{19}=(0.0,0.97,-1.14)^{\top}$, would  desynchronize the network. The blue color corresponds to time intervals for which the perturbation restores synchronization. (b) A highlight of the squared region of (a). The inset shows the sensitivity to the perturbation timing at a finer scale.}
\label{figure_2}
\end{figure}

To clarify this matter, we check the response to a particular perturbation applied at different states along the trajectory on the completely synchronized limit cycle ${\bf A_S}$. First, we define an order parameter $\mathcal{Z}= 1/N \sum_{i=1}^{N}K_i$ with $K_i=0$ for $E_i(t_{end})<\delta$, and $K_i=1$ for $E_i(t_{end})>\delta$. The overall integration time, $t_{end}$, is fixed at $t_{end}=3 \times 10^4$. Employing this definition, the completely synchronized state gives $\mathcal{Z}=0$, while the completely desynchronized state gives $\mathcal{Z}=1$. The parameter $\delta=0.01$ controls the synchronization quality. In Fig. \ref{figure_2}(a), we show the component $z_S$ of the synchronized oscillatory state as a function of the perturbation time $t_p$. The blue colored points indicate the instants of time, in which the perturbation ${\bf \Delta}_{19}=(0.0,0.97,-1.14)^{\top}$ leads to restoration of synchronization in the network, $\mathcal{Z}=0$. The red colored points indicate the instants $t_p$ in which the same perturbation, ${\bf \Delta}_{19}$, would desynchronize the network, $\mathcal{Z}=1$. In Fig. \ref{figure_2}(b), we highlight a very sensitive time interval of the synchronized oscillation (squared region of Fig. \ref{figure_2}(a)). In the inset, we show a magnification of a time interval in which both outcomes are mixed even at a finer scale. This confirms the observations of Fig. \ref{figure_1}, and shows that the distribution of time instants leading to synchronization or desynchronization is very intricate, exhibiting a fractal-like behavior with more and more mixing of outcomes at finer and finer scales.       
 
\begin{figure*}[!htp] 
\centering
\includegraphics[width=18cm,height=9cm]{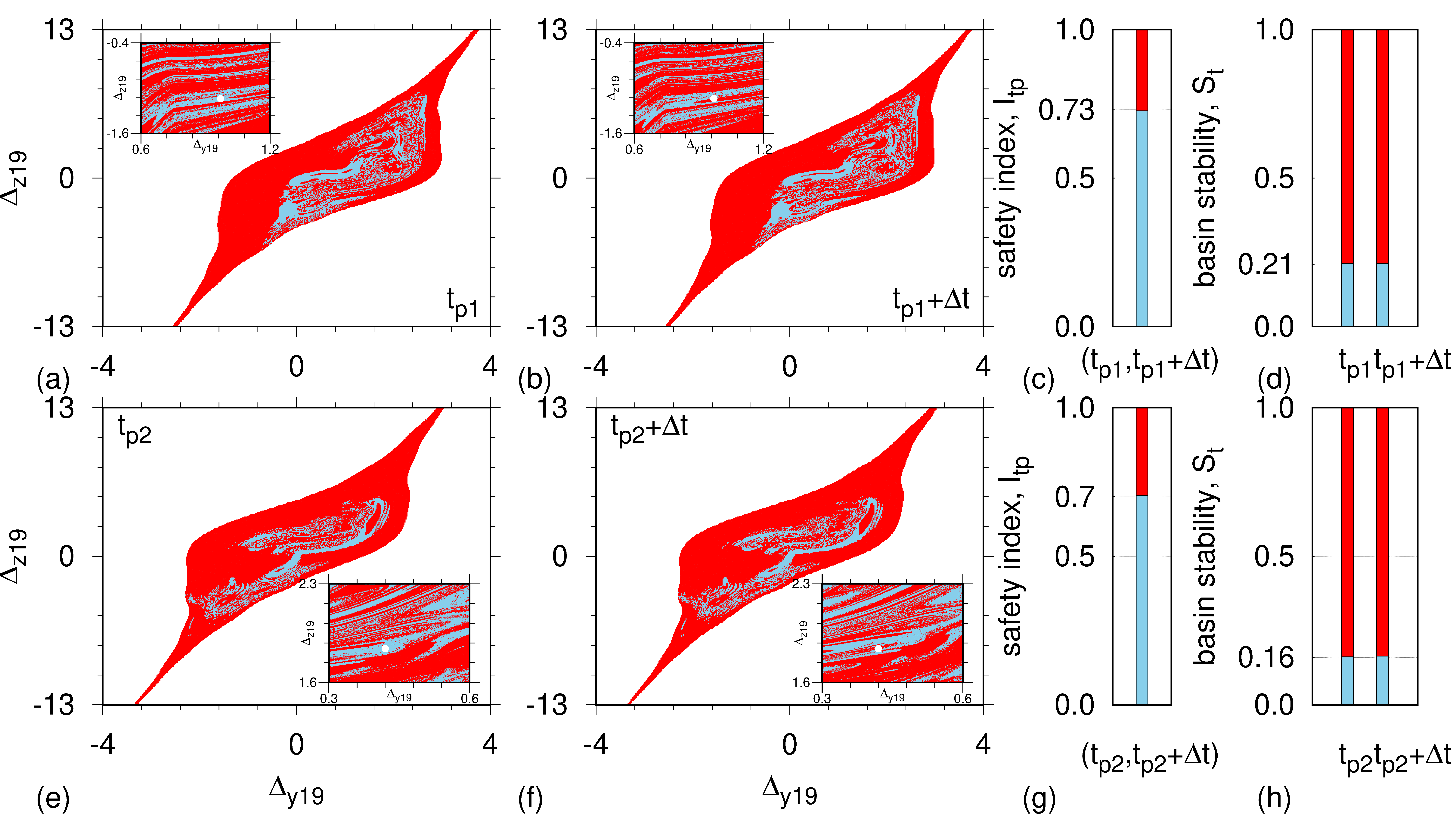}
\caption{(a) and (b) Synchronization diagrams, $\Delta z_{19} \times \Delta y_{19}$, of perturbations amplitude for instants of time $t_{p1}=4005.97$ and $t_{p1}+\Delta t=4005.98$, respectively. The color codifies the order parameter, blue for $\mathcal{Z}=0$ (synchronized) and red for $\mathcal{Z}=1$ (completely desynchronized). In (c), the blue color indicates the safety index between (a) and (b). In (d) the blue color indicates the basin stability of (a) and (b). (e)-(h) The same procedure for two different consecutive instants of time $t_{p2}=4009.95$ and $t_{p2}+\Delta t=4009.96$. The white dots in the insets indicate changes of $\mathcal{Z}$}
\label{figure_3}
\end{figure*}

Next, we investigate the synchronization dynamics in the space of perturbations ${\bf \Delta}_i$ for fixed instants of time. For the sake of visualization, we restrict ourselves to applying perturbations in the plane $\Delta x_i=0$. In Fig. \ref{figure_3}(a), for the perturbation time instant fixed at $t_{p1}=4005.97$, we show a synchronization diagram for $(\Delta z_{19} \times \Delta y_{19})$. The blue color indicates perturbation regions for which the network synchronizes ($\mathcal{Z}=0$), the safe sets. The red color indicates regions for which the network desynchronizes ($\mathcal{Z}=1$). Regions in white indicate perturbations for which the solution converges to infinity, the second attractor in the system. In Fig. \ref{figure_3}(a), we find continuous regions of perturbations leading the network to both, synchronized or desynchronized, states. Additionally, we also observe regions where the perturbations leading to each behavior are very intricate even resembling a riddled-like synchronization basin \cite{Medeiros2018}. In Fig. \ref{figure_3}(b), we present the synchronization diagram for a subsequent instant of time, $t_{p1}+\Delta t=4005.98$, and obtain similar characteristics for the distribution of the blue points. However, if the procedure is repeated for other instants of time $t_{p2}=4009.95$ and $t_{p2}+\Delta t=4009.96$ [Fig. \ref{figure_3}(e) and \ref{figure_3}(f)], we find a completely different distribution of such points, though the pictures of subsequent instants of time are again similar. 

In order to determine the changes between perturbation planes obtained for subsequent instants of time causing the sensitivity to timing, we define a finite perturbation plane as $U = \{ \Delta_{19} \in \mathbb{R}^3 \mid \Delta x_{19}=0, \Delta y_{19} \in [-4,4], \Delta z_{19} \in [-13,13] \}$, as in the diagrams shown in Fig. \ref{figure_3}. The safe set at the time instant $t_p$ is defined as the subset of $U$ for which $\mathcal{Z}=0$, and denoted by $\mathcal{B}^{\mathbf{S}}_{t_p} \subset \mathcal{B}(\mathbf{A_S})$. Similarly, the unsafe sets, the perturbations in $U$ that desynchronize the network, are defined as the elements of $U$ for which $\mathcal{Z}>0$, and denoted by $\mathcal{B}^{{\bf \Lambda}}_{t_p}$. Now, we estimate the fraction of the safe set $\mathcal{B}^{\mathbf{S}}_{t_p}$ that synchronizes the network at the time instant $t_p$ and also at $t_p+\Delta t$ by:
\begin{equation}
 I_{t_p}=Vol(\mathcal{B}_{t_p}^{\mathbf{S}} \cap \mathcal{B}_{t_p+\Delta t}^{\mathbf{S}})/Vol(\mathcal{B}_{t_p}^{\mathbf{S}}). 
\end{equation}
We call this measure the safety index, as it reflects the probability of the network to possess the same response with respect to perturbations at subsequent time instants. Hence, in Fig. \ref{figure_3}(c), we find $I_{t_{p1}}=0.73$, as the safety index for perturbations applied at $t_{p1}=4005.97$ and $t_{p1}+\Delta t=4005.98$, this indicates that only $73\%$ of the perturbations at $t_{p1}$ still synchronize the network at the instant $t_{p1}+\Delta t$. The same procedure is applied to the perturbation time instants at $t_{p2}=4009.95$ and $t_{p2}+\Delta t=4009.96$ resulting in a safety index $I_{t_{p2}}=0.70$, Fig. \ref{figure_3}(g). The white dots in the insets of Fig. \ref{figure_3} indicate changes, from $0$ to $1$, of the order parameter for subsequent instants of time. These results indicate that the safe set $\mathcal{B}_{t_p}^{\mathbf{S}}$ changes for every instant of time, causing the system to be sensitive to the timing of perturbations with the same amplitude. Increasing the time difference $\Delta t$, the safety index would decrease accordingly. Comparing the safe sets of Figs. \ref{figure_3}(a) and (e), where $\Delta t=t_{p2}-t_{p1}\approx 4.0$, the dissimilarity is evident. To investigate whether the real-time vulnerability is related with the relative size of the safe sets, we compute their {\it basin stability} \cite{Menck2013}. To this end, we first denote the subset of perturbations for which only finite solutions are observed as $\mathcal{Q}_{t_p}=\mathcal{B}_{t_p}^{\mathbf{S}} \cup \mathcal{B}_{t_p}^{\bf \Lambda}$, i.e., all perturbations leading to infinity are excluded. Then, we estimate the measure $S_{t_p}=Vol(\mathcal{B}_{t_p}^{\mathbf{S}})/Vol(\mathcal{Q}_{t_p})$ that constitutes an estimate of the volume of $\mathcal{Q}_{tp}$ occupied by the safe set $\mathcal{B}^{\mathbf{S}}_{t_p}$. For the synchronization diagrams, we obtain  $S_{tp1}=S_{tp1+\Delta t}=0.213$ [Fig. \ref{figure_3}(d)] and $S_{tp2}=0.161$ and $S_{tp2+\Delta t}=0.164$ [Fig. \ref{figure_3}(h)]. These findings demonstrate that the relative volume of the safe sets does not change significantly for subsequent instants of time, i.e., $S_{t_p} \approx S_{t_p+\Delta t}$. Therefore, real time vulnerability is only related to changes in the location of $\mathcal{B}^{\mathbf{S}}_{t_p}$ with respect to each point on the limit cycle ${\bf A_S}$ and not to the relative size of the safe sets. As a consequence, the safety index is a suitable indicator of this kind of vulnerability, while basin stability is not a sensitive measure. Next, we discuss the mechanism causing the intricate dependence of synchronization on the perturbation time instant $t_p$.  

\begin{figure}[!htp] 
\centering
\includegraphics[width=8.5cm,height=3.5cm]{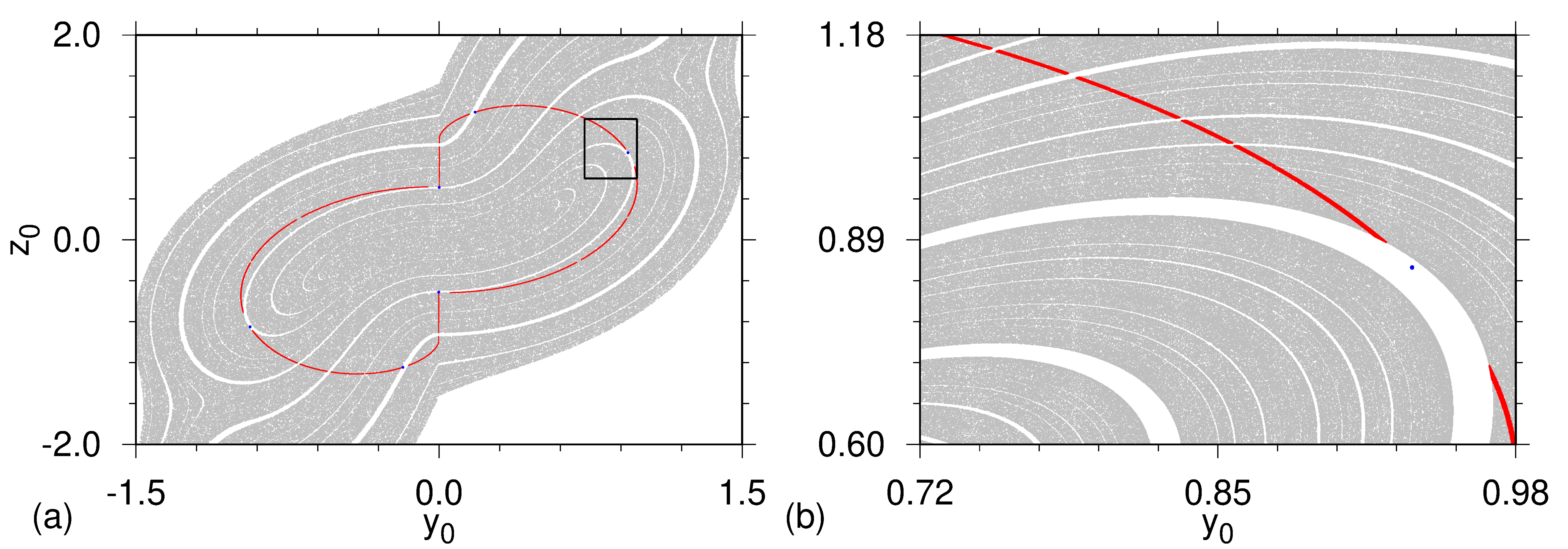}
\caption{(a) Basin of attraction of the limit cycle $\mathbf{A}$ on the plane $x_0=0$. The six blue dots represent the intersections of $\mathbf{A}$ with this section, while red dots mark the projection of the chaotic saddle $\Lambda$. Gray dots mark the initial conditions having lifetime longer than $50$ and provide an approximation to the stable manifold of $\Lambda$. (b) A magnication of the black ractangle in (a) showing the detail of the chaotic saddle. Note how close the saddle $\Lambda$ falls to $\mathbf{A}$.}
\label{figure_4}
\end{figure}

The mechanism behind this phenomenon is related to the chaotic set $\mathbf{\Lambda}$ lying very close to the stable limit cycle $\mathbf{A_S}$. This high-dimensional chaotic set $\mathbf{\Lambda}$ appears from the individual ones, occurring in every single circuit. Considering a Poincar\'e section at $x=0$ for an individual uncoupled circuit, we show in Fig. \ref{figure_4} a projection of the chaotic saddle $\Lambda$ (red dots) obtained by the sprinkler method \cite{Tamas2011} and an approximation of its stable manifold (gray dots). The escape time is defined as the number of crossings in the Poincare section before a disk of radius $\varepsilon=10^{-4}$ around the attractor points is reached, and obeys an exponential distribution \cite{Tamas2011,Lilienkamp2017}. However, coupling all those circuits, another large chaotic set $\mathbf{\Lambda}$ emerges being difficult to obtain \cite{Sweet2001}. As demonstrated in \cite{Medeiros2018}, this chaotic set appears to be an attractor for the coupled system or, at least, a chaotic saddle with extremely long transients with escape times beyond numerical computations. Hence, when the system is coupled, there is a competition, in each unit, between the network coupling and the chaotic dynamics in the vicinity of the chaotic set $\mathbf{\Lambda}$. If the coupling is not strong enough to attract the perturbed unit fastly back to the synchronization manifold, despite the perturbed unit, additional ones are pulled to the chaotic set. Once a critical number of units approach this chaotic set, escaping from it become very unlikely \cite{Crutchfield1988,Lai1995}, trapping the high-dimensional system in a chaotic desynchronized behavior for times indefinitely long. This mechanism is explained in detail in \cite{Medeiros2018} for perturbations applied in the initial instant of time. Here, we show a much stronger consequence of such a phenomenon, the outcome of such a competition between the coupling strength and the chaotic dynamics leading to synchronization or desynchronization, exhibits an intricate dependence on the timing of the perturbations.

\begin{figure}[!htp] 
\centering
\includegraphics[width=8cm,height=5.5cm]{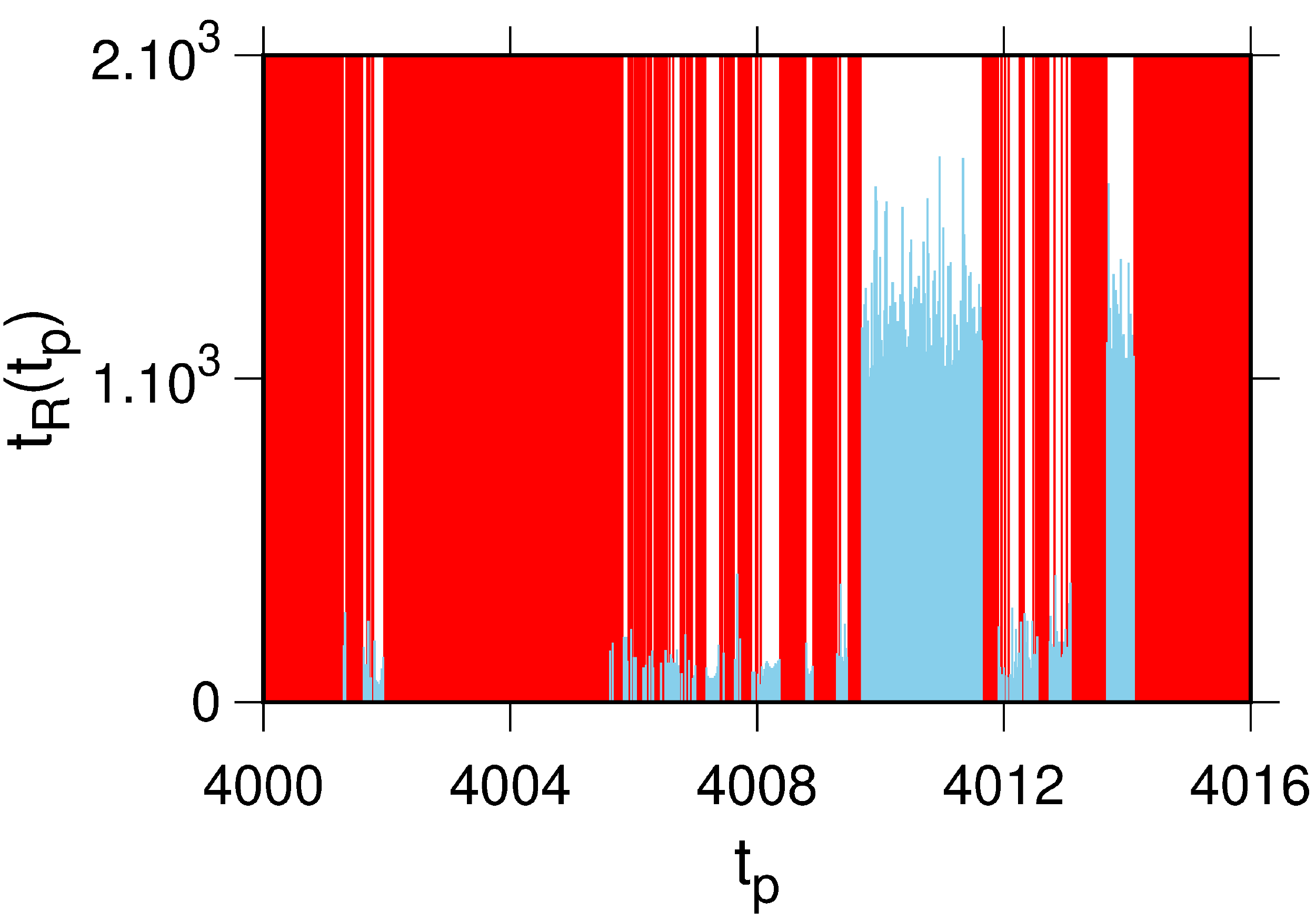}
\caption{Time for the synchronized state be restored, $t_R$, as function of the perturbation instant, $t_p$. The perturbation is given by ${\bf \Delta}_{19}=(0.0,0.97,-1.14)^{\top}$. The red bars delimitate instants of time for which the system desynchronizes, while the height of the blue bars represents the restoring time of synchronization.}
\label{figure_5}
\end{figure}

Now, in order to demonstrate the time dependence of the phenomenon, we analyze the trajectory of the perturbed unit for perturbations, ${\bf \Delta}_{19}=(0.0,0.97,-1.14)$, applied at different instants of time. For the instants of time at which the synchronization manifold is restored, the system can be examined by computing the transient time to return to the synchronized state \cite{Lilienkamp2018}. In Fig. \ref{figure_5}, we show the return time $t_R$, needed to restore complete synchronization as a function of the time instant $t_p$, in which the perturbation is applied. The red bars correspond to instants of time for which full desynchronization occurs, i.e., $t_R \rightarrow \infty$. The height of the blue bars indicates the finite return times for perturbations applied at their respective instant of time. The variability found in the finite values of $t_R$ and the nontrivial distribution of the red bars indicates the sensitivity encountered by the perturbed unit depending on the perturbation timing.

We report the existence of an intricate time dependence of the vulnerability of the synchronized states in a network composed of identical electronic circuits. By perturbing the synchronized dynamics in consecutive instants of time, we find that synchronization breaks down for some time instants while it persists for others. The mechanism behind this intriguing phenomenon is the existence of a chaotic set close to the synchronized trajectory. Such saddle is a rather common phenomenon and can be found in any dynamical system which possesses parameter ranges in which chaotic dynamics are interspersed with periodic windows. Besides the periodic synchronization discussed here, the same phenomenon may also occur for systems synchronized in a chaotic attractor. Therefore, real-time vulnerability of synchronized states is a very ubiquitous phenomenon. Apart from warning about vulnerabilities along synchronized trajectories, this phenomenon offers to determine a time window for the success of interventions aiming to break synchronized behavior in complex systems. For instance, a hypersynchronous state in the brain must be suppressed in order to terminate an epileptic seizure.

This work was supported by FAPESP (Processes: 2018/03211-6, 2013/26598-0, 2015/50122-0, 2017/05521-0) and the National Research, Development and Innovation Office of Hungary under grant K-125171.

\end{document}